\documentclass{article}

\usepackage[utf8]{inputenc} 
\usepackage[T1]{fontenc}    
\usepackage{hyperref}       
\usepackage{url}            
\usepackage{fancyhdr}       
\usepackage{hyphenat}       
\usepackage{geometry}       

\title{A Phenomenological Approach to Analyzing User Queries in IT Systems Using Heidegger’s Fundamental Ontology}
\author{
  Maksim Vishnevskiy \\
  Seminars "Philosophy + IT" \\
  \href{mailto:vmaxims@gmail.com}{vmaxims@gmail.com}
}

\begin{document}

\maketitle

\begin{abstract}
This paper presents a novel research analytical IT system grounded in Martin Heidegger’s \emph{Fundamental Ontology}, distinguishing between \emph{beings} (das Seiende) and \emph{Being} (das Sein). The system employs two modally distinct, descriptively complete languages: a categorical language of beings for processing user inputs and an existential language of Being for internal analysis. These languages are bridged via a phenomenological reduction module, enabling the system to analyze user queries (including questions, answers, and dialogues among IT specialists), identify recursive and self-referential structures, and provide actionable insights in categorical terms. Unlike contemporary systems limited to categorical analysis, this approach leverages Heidegger’s phenomenological existential analysis to uncover deeper ontological patterns in query processing, aiding in resolving logical traps in complex interactions, such as metaphor usage in IT contexts. The path to full realization involves formalizing the language of Being by a research team based on Heidegger’s \emph{Fundamental Ontology}; given the existing completeness of the language of beings, this reduces the system’s computability to completeness, paving the way for a universal query analysis tool. The paper presents the system’s architecture, operational principles, technical implementation, use cases—including a case based on real IT specialist dialogues—comparative evaluation with existing tools, and its advantages and limitations.
\end{abstract}


\section{Introduction}
Heidegger’s philosophy, though primarily ontological, has found selective but significant applications in technology and computer science, offering frameworks to rethink human-technology interactions. Dreyfus (1991) applied Heidegger’s concept of \emph{being-in-the-world} to critique early AI paradigms, highlighting their neglect of contextual engagement. Winograd and Flores (1986) drew on Heidegger to reconceptualize system design, emphasizing context and meaning in human-computer interaction. Dourish (2001) advanced these ideas in human-computer interaction (HCI), advocating phenomenological approaches to interface design. In natural language processing (NLP), technical advancements like Transformer models have enabled context-sensitive query processing, yet philosophical perspectives remain underexplored. This paper builds on these foundations, applying Heidegger’s \emph{Fundamental Ontology} to enhance user query analysis in IT systems, particularly in dialogues involving metaphorical and technical descriptions.

The growing complexity of user queries in IT systems, especially in artificial intelligence research and particularly artificial general intelligence (AGI), demands tools capable of identifying and resolving logical inconsistencies, such as recursive patterns arising from self-reference in questions and answers. Traditional analytical systems like IBM Watson or Palantir excel at categorical analysis but fail to address ontological issues tied to deeper thought structures. This paper describes an IT system concept based on Heidegger’s \emph{Fundamental Ontology}, distinguishing \emph{beings} and \emph{Being}. The system uses two modally distinct, non-overlapping languages: categorical for user inputs and existential for internal analysis, linked via phenomenological reduction. This enables analysis of user queries, including dialogues, identifying recursions, and providing solutions in accessible categorical terms. Full realization requires formalizing the language of Being by a research team focused on its structure within Heidegger’s \emph{Fundamental Ontology}; given the descriptive completeness of the language of beings, this ensures the system’s computability at the level of completeness, making it a universal tool for ontological query analysis.

The system does not aim to provide operational definitions of concepts like \emph{consciousness} or \emph{Being} for direct implementation in programming language syntax (e.g., C++ or Prolog), as is common in traditional cybernetics. Instead, \emph{beings} and \emph{Being} serve as distinct, non-mixing environments structuring the system’s architecture, designed to analyze recursive patterns in user queries—e.g., in NLP tasks for AGI, where traditional tools detect logical loops (e.g., “consciousness defines subjectivity, and subjectivity defines consciousness”) but offer no resolutions.

The system is relevant for processing queries in interdisciplinary scientific contexts, where programmers, physicists, mathematicians, neuroscientists, philosophers, and other experts explore AGI’s subjectivity and consciousness. The paper presents the system’s principles (Section 2), internal structure (Section 3), user interface (Section 4), technical implementation (Section 5), use cases based on user queries (Section 6), comparison with modern systems (Section 7), and discussion of significance and limitations (Section 8).

\section{Main Principles}
The system is grounded in Heidegger’s distinction between \emph{beings} (das Seiende) and \emph{Being} (das Sein), implemented through two modally distinct, non-overlapping languages:
\begin{itemize}
    \item \textbf{Language of beings (das Seiende):} A categorical language describing entities, their properties, and relations (e.g., “query terms,” “user intent,” “computational processes” in IT contexts, or “concepts,” “universality,” “experience” in conceptual query analysis). Used for processing user queries.
    \item \textbf{Language of Being (das Sein):} An existential language describing modes and modalities of Being through concepts like Dasein, openness, non-spatio-temporal being, temporality, and others. Applied for internal ontological analysis of query structures.
\end{itemize}
These languages remain separated due to their modal differences, with interaction facilitated by a phenomenological reduction module linking categorical and existential interpretations. Notably, the “language of beings” and “language of Being” are not traditional programming constructs (e.g., C++ procedures) but structural environments defining the system’s architecture. The categorical language of beings is implemented using existing technologies like RDF/OWL and graph models, while the existential language of Being draws on the modes and modalities of Heidegger’s \emph{Fundamental Ontology}, such as Dasein, openness, and temporality.

\section{Internal System Operation}
\subsection{User Query Analysis}
The system processes user queries (e.g., questions and answers in dialogues) using NLP methods, capturing phenomena at a pre-categorical level. For instance, the question “Who are we?” in an AGI-related query or the statement “Some concepts are universal” in an experience-related query is treated as “something” without immediate categorization.

\subsection{Dual Interpretation}
Each phenomenon is interpreted simultaneously in both languages:
\begin{itemize}
    \item \textbf{Categorical description (language of beings):} The phenomenon is broken into categories like “query terms,” “user intent,” or “cognitive processes” for IT queries, or “concepts,” “universality,” “orientation,” and “culture” for conceptual queries, structured as ontological graphs.
    \item \textbf{Existential description (language of Being):} The phenomenon is analyzed through \emph{Fundamental Ontology} structures, e.g., as a mode of Dasein tied to openness to the world and the existential of “care.”
\end{itemize}
The phenomenological reduction module links these interpretations, providing a dual perspective on the phenomenon.

\subsection{Recursion Detection}
The system identifies recursive structures using:
\begin{itemize}
    \item \textbf{Graph models:} Queries are represented as graphs, with nodes as assertions and edges as logical connections. Cycles indicate recursion, e.g., “consciousness” via “subjectivity” and back in AGI queries, or “universality of concepts” via “experience” and back in conceptual queries.
    \item \textbf{Semantic analysis:} Detects self-referential constructs, such as defining “consciousness” through “subjectivity” and “subjectivity” through “consciousness,” or “concepts” through “subject’s experience” and “experience” through “concepts.”
\end{itemize}

\subsection{Query Recursion in Practice}
In real-world IT query processing, recursive patterns often arise when users focus on isolated query fragments rather than holistic ideas. For example, in Transformer-based model queries, a user may misinterpret “dynamic patterns” as implying a lack of structure, leading to a cycle where responses address the fragment (e.g., defending model architecture) instead of the intended concept (e.g., distinguishing data from interpretation). Such recursions reflect logical loops in query processing, where assertions cycle without resolution. The proposed system addresses this by separating data (query fragments) and meaning (intended ideas) through phenomenological reduction, identifying cycles and offering solutions in categorical terms, enhancing clarity in interdisciplinary IT contexts like NLP and ontology management.

\subsection{Analytical Output}
Upon detecting recursion, the reduction module generates:
\begin{itemize}
    \item Identification of the recursive pattern in categorical terms.
    \item Suggestions for resolution using the language of beings.
\end{itemize}

\section{User Interface}
The system’s user interface employs only the language of beings, presenting outputs as text, graphs, or interactive elements. Examples include:
\begin{itemize}
    \item \textbf{Text:} “Detected recursion: ‘we’ is tied to ‘subjectivity,’ which depends on ‘we.’ Consider ‘we’ through ‘cognitive structures’ or ‘social processes’ instead of ‘subjectivity.’” Or: “Detected recursion: ‘universality of concepts’ depends on ‘experience,’ which depends on ‘concepts.’ Consider ‘universality’ through ‘cultural systems.’”
    \item \textbf{Graph:} Visualization of the query cycle with a suggested break point.
\end{itemize}

\section{Technical Implementation}
\subsection{Contextual Analysis}
NLP extracts pre-categorical phenomena from user queries, representing them as vector embeddings for queries about AGI subjectivity, concept universality, etc.

\subsection{Dual Language Interpretation}
\begin{itemize}
    \item \textbf{Language of beings:} Implemented via ontologies (OWL, RDF), structuring phenomena as entities (e.g., “user intent,” “world model” for IT queries; “universality,” “experience” for concepts). Can be coded in Python or C++ using graphs and semantic analysis.
    \item \textbf{Language of Being:} Implemented through formal \emph{Fundamental Ontology} structures, describing phenomena as modes and modalities of Dasein. Relies on Heidegger’s vocabulary (Dasein, care, non-spatio-temporal being, understanding, openness), rich in semantics but existing in a modality distinct from traditional programming languages. Its structures await formalization for computational use.
\end{itemize}

\subsection{Phenomenological Reduction}
The system is designed to avoid recursion caused by self-reference, with phenomenological reduction playing a pivotal role. If the language of Being is viewed as an architectural layer of formal structures (the whole) and the language of beings as a layer of concrete structures (instances), the phenomenological reduction module functions not merely as a tool to eliminate recursions, akin to a boundary between non-overlapping layers. It enables conscious management of semantic patterns within the whole’s structures.

The module operates by parallelizing two simultaneous, non-mixing interpretations, transferring semantic patterns from instance structures to whole structures—similar to how a parser and optimizer in an 80386 processor manage data flows without creating loops. This process prevents semantic pattern cycling while organizing and optimizing layer interactions.

Thus, phenomenological reduction enables the system to analyze query phenomena, avoiding recursive traps, and provides a toolkit for conscious meaning transfer and transformation between the two architectural levels—language of Being and language of beings.

\subsection{Recursion Detection and Output}
Graph algorithms and semantic analysis detect recursions in queries, with outputs translated into the language of beings, offering solutions for IT and conceptual queries.

\section{Use Cases}
\subsection{Scenario 1: AGI Subjectivity}
\textbf{Scenario:} In a query submitted to an AGI-research IT system, a user asks: “Can AGI possess subjectivity if we provide it a world model?” A clarifying question arises: “Who are ‘we’ in this context?” Another user responds: “‘We’ is a system with subjectivity that creates and transfers the world model to AGI.”

\textbf{Analysis:}
\begin{itemize}
    \item \textbf{Pre-categorical perception:} The query “Who are we?” is captured as “something.”
    \item \textbf{Dual interpretation:}
        \begin{itemize}
            \item \textbf{Categorical:} “We” is linked to “subjectivity,” “world model,” “cognitive processes.”
            \item \textbf{Existential:} “We” as a mode of Dasein, openness to the world via the existential of “care.”
        \end{itemize}
    \item \textbf{Recursion detection:} A cycle emerges: “we” is defined through “subjectivity,” and “subjectivity” through “we.”
    \item \textbf{Phenomenological reduction:} The existential perspective suggests reinterpreting “we” as a structure of Being, not merely a categorical entity.
\end{itemize}

\textbf{Output:} “Detected recursion: ‘we’ is tied to ‘subjectivity,’ which depends on ‘we.’ To avoid recursion, consider ‘we’ through ‘cognitive structures’ or ‘social processes’ instead of ‘subjectivity.’”

\subsection{Scenario 2: Universality of Concepts and Their Relation to Experience}
\textbf{Scenario:} In a query submitted to an IT system, a user asks: “Are concepts like ‘earth–sky’ or ‘right–left’ universal across humans due to fundamental bodily experience?” Another user objects: “It’s not universal, as some cultures use directions like ‘north–south,’ e.g., in languages saying ‘object southwest of me.’” The first user clarifies that bodily experience (e.g., hand differentiation) may still underpin universality, but the second insists this experience is culturally conditioned.

\textbf{Analysis:}
\begin{itemize}
    \item \textbf{Pre-categorical perception:} The query “Are concepts universal?” is captured as “something.”
    \item \textbf{Dual interpretation:}
        \begin{itemize}
            \item \textbf{Categorical:} “Universality” is linked to “concepts,” “experience,” “orientation,” “body,” “hands,” “cardinal directions,” “culture.”
            \item \textbf{Existential:} “Universality” as a mode of Dasein, disclosing its Being through world interaction, yet inevitably tied to categorical interpretations.
        \end{itemize}
    \item \textbf{Recursion detection:} A cycle emerges: “universality of concepts” is defined through the subject (whose experience shapes concepts), and the subject through “concepts” (ability to perceive and interpret).
    \item \textbf{Phenomenological reduction:} The existential perspective suggests viewing “universality” as a modal structure of Being, not a fixed category.
\end{itemize}

\textbf{Output:} “Detected recursion: ‘universality of concepts’ is defined through ‘experience,’ which depends on the subject, and the subject through ‘concepts.’ To resolve, consider ‘universality’ as a context-dependent category, e.g., through ‘physiological processes’ or ‘cultural systems,’ rather than an absolute property.”

\subsection{Scenario 3: Defining Consciousness in Cybernetics}
\textbf{Scenario:} In a query about applying technical cybernetics to ontology, a user asserts: “No scientifically necessary definition of ‘consciousness’ exists in natural language, so it’s irrelevant to technical cybernetic languages like C++ or Prolog until defined in their base constructs.” Another user objects: “But this demand for definition loops—how to define ‘consciousness’ if it defines itself, as in ‘consciousness understands itself’?” The first user insists that without definitions, the query lacks practical meaning.

\textbf{Analysis:}
\begin{itemize}
    \item \textbf{Pre-categorical perception:} The query “What is consciousness?” is captured as “something.”
    \item \textbf{Dual interpretation:}
        \begin{itemize}
            \item \textbf{Categorical:} “Consciousness” is linked to “definition,” “understanding,” “technical constructs,” “cybernetics.”
            \item \textbf{Existential:} “Consciousness” as a mode of Dasein, tied to self-understanding and being-in-the-world.
        \end{itemize}
    \item \textbf{Recursion detection:} A cycle emerges: “consciousness” requires “definition” for implementation, but “definition” depends on understanding “consciousness,” looping back to self-reference.
    \item \textbf{Phenomenological reduction:} The existential perspective suggests reinterpreting “consciousness” as a structure of Being of Dasein, not a static categorical definition.
\end{itemize}

\textbf{Output:} “Detected recursion: ‘consciousness’ is tied to ‘definition,’ which depends on ‘understanding consciousness,’ looping back to ‘consciousness.’ To resolve, consider ‘consciousness’ through ‘functional processes’ or ‘interactive systems’ instead of a fixed definition.”

\subsection{Scenario 4: Metaphors in Generative AI Queries}
\textbf{Scenario:} In a query about generative AI (GenAI), a user describes Transformer models as an “artificial neocortex,” implying brain-like processing. Another user challenges the term, arguing it obscures technical details, leading to a cycle: the first defends the metaphor as intuitive, while the second insists on categorical descriptions, but neither clarifies the model’s functions.

\textbf{Analysis:}
\begin{itemize}
    \item \textbf{Pre-categorical perception:} The query term “artificial neocortex” is captured as “something.”
    \item \textbf{Dual interpretation:}
        \begin{itemize}
            \item \textbf{Categorical:} “Neocortex” is linked to “Transformer,” “processing,” “architecture.”
            \item \textbf{Existential:} “Neocortex” as a mode of disclosing system behavior, tied to human-like understanding.
        \end{itemize}
    \item \textbf{Recursion detection:} A cycle emerges: “neocortex” is defined through “Transformer” (as brain-like), and “Transformer” through “neocortex” (as intuitive), looping without technical clarity.
    \item \textbf{Phenomenological reduction:} The existential perspective reinterprets “neocortex” as an interpretive layer, not a technical structure.
\end{itemize}

\textbf{Output:} “Detected recursion: ‘neocortex’ is tied to ‘Transformer,’ which loops back to ‘neocortex.’ To resolve, consider ‘Transformer’ through ‘functional processes’ or ‘data flows’ instead of metaphorical terms.”

\subsection{Scenario 5: Metaphors in IT Communication}
\textbf{Scenario:} In a dialogue with an IT specialist, a user discusses using metaphors like “neocortex” or “neuron” to describe generative AI (GenAI) systems, suggesting they inspire design ideas, e.g., for code refactoring or module structuring. The specialist notes that such metaphors are used informally (“over tea”) or in documentation to aid understanding but acknowledges “things can get muddled.” For example, describing a module as a “neuron” may spark creativity but risks misinterpretation by team members, potentially delaying project clarity. The specialist values metaphors, likening them to “poetry” requiring effort to yield results, unlike “fiction”—superficial rhetoric—but paradoxically uses poetic imagery to describe their practical value. He asserts that thinking occurs through images, not words, citing Einstein’s visual and motor imagery to justify metaphors as fundamental to cognition, and mentioning Buddhist concepts (“karma” and “skandhas”) to frame images as shaping lived reality. He insists that metaphors in industrial development, embedded in functional system design (FSD) and project documentation, pose no risk if developers are logical, confidently claiming errors are unlikely due to robust logic and thorough impact analysis in experienced individual development. He dismisses concerns about potential confusion as “odd,” reflecting overconfidence in his approach. He demands “real systems”—tangible tools—over conceptual discussions, rejecting the latter as irrelevant to his operational semantics (experience of a specific group of specialists) and expressing skepticism about the formalizability of phenomenological reduction due to its individual nature. He misinterprets queries about joining his operational semantics (experience of a specific group) as invitations to engage with theoretical concepts, demonstrating a logical conflation that excludes external perspectives, underscoring a pragmatic and cognitive barrier to ontological analysis.

\textbf{Analysis:}
\begin{itemize}
    \item \textbf{Pre-categorical perception:} The query term “metaphor” (e.g., “neuron” or “neocortex”) is captured as “something.”
    \item \textbf{Dual interpretation:}
        \begin{itemize}
            \item \textbf{Categorical:} “Metaphor” is linked to “module,” “code,” “design,” “team communication,” “documentation.”
            \item \textbf{Existential:} “Metaphor” as a mode of Dasein’s disclosure, revealing system behavior through creative interpretation but risking ambiguity in shared understanding.
        \end{itemize}
    \item \textbf{Recursion detection:} A cycle emerges: “metaphor” (e.g., “neuron”) is used to describe “module design,” but “module design” is understood through “metaphor,” looping when team members interpret the metaphor differently, delaying clarity.
    \item \textbf{Phenomenological reduction:} The existential perspective reinterprets “metaphor” as an interpretive layer distinct from technical structures, suggesting a categorical alternative to break the cycle.
\end{itemize}

\textbf{Output:} “Detected recursion: ‘metaphor’ (e.g., ‘neuron’) is tied to ‘module design,’ which loops back to ‘metaphor’ due to differing interpretations. To resolve, consider ‘module design’ through ‘functional requirements’ or ‘data structures’ instead of metaphorical terms, ensuring team alignment.”

\section{Comparison with Modern Systems}
\subsection{Modern Systems}
Tools like IBM Watson:
\begin{itemize}
    \item Operate solely in the language of beings.
    \item Detect cycles (e.g., in AGI subjectivity, concept universality, consciousness definitions, or metaphorical IT dialogues) but offer no ontological resolutions.
    \item Are limited to descriptive outputs.
\end{itemize}

\subsection{Advantages of the Proposed System}
\begin{itemize}
    \item \textbf{Existential depth:} The hidden language of Being provides ontological analysis for IT queries, concepts, cybernetic definitions, and specialist dialogues.
    \item \textbf{Practical outputs:} Recommendations overcome recursions in all cases. The system’s advantage lies in not only detecting recursions (e.g., “consciousness defines subjectivity, and subjectivity defines consciousness” in IT queries or “metaphor defines design, and design defines metaphor” in IT communication) as modern tools do but also offering ontologically grounded resolutions via existential analysis, particularly valuable for NLP in AGI contexts and team collaboration.
    \item \textbf{Universality:} Pre-categorical perception applies to any user query, from AGI to IT specialist dialogues.
    \item \textbf{Compatibility with non-linear methods:} Unlike traditional systems focused on linear categorical models, the proposed system aligns with modern non-linear approaches, such as those inspired by computational irresolvability, analyzing query patterns as dynamic phenomena. The phenomenological reduction module enhances GenAI applications, resolving recursive misunderstandings without relying on static ontologies, bridging existential analysis and practical NLP tasks.
\end{itemize}

\subsection{Limitations}
\begin{itemize}
    \item \textbf{Implementation complexity:} Formalizing the language of Being and phenomenological reduction module structures requires significant organizational and managerial resources to establish and coordinate a research team capable of focusing cognitive and intellectual resources on developing a necessary and sufficient description of the language of Being’s structure based on Heidegger’s \emph{Fundamental Ontology}. This is a labor-intensive process dependent on systems engineers and researchers with deep phenomenological knowledge and specialized expertise. Specialists often express skepticism about such formalization, viewing ontological processes as too individual for system integration, exacerbated by cognitive biases like overconfidence in their logic and processes, hindering external perspective adoption. Secondarily, implementation depends on computational power and memory volume needed to process formalized structures.
    \item \textbf{Limited visibility:} The system’s design prioritizes practical outputs in the language of beings over teaching users the language of Being, so the existential language remains internal.
\end{itemize}

\section{Discussion and Conclusions}
Through its phenomenological reduction structure, the system integrates Martin Heidegger’s existential analysis with computational analysis, enabling it to overcome recursive traps of self-reference in user queries. Its modally distinct bilingual structure and phenomenological reduction module provide analytical depth surpassing modern tools. Diverse queries across the scenarios demonstrate the system’s versatility in resolving recursions arising from self-reference and metaphorical ambiguity.

In Scenario 1, the AGI subjectivity query revealed a cycle between “we” and “subjectivity,” where defining one through the other created recursion. The system suggested reformulating “we” through cognitive or social processes, avoiding self-reference. Similarly, in Scenario 2, the discussion of concept universality (“earth–sky,” “right–left”) showed a recursion between “concepts” and “experience,” with cultural and physiological interpretations looping. The system resolved this by suggesting context-dependent categories like physiology or culture. Scenario 3, addressing consciousness definition in cybernetics, identified a cycle between “consciousness” and its “definition,” where demanding a precise definition led to self-reference. The system proposed functional or interactive approaches to break the cycle. In Scenario 4, the “artificial neocortex” metaphor for Transformer models created a recursion, as intuitive description looped with technical aspects, and the system recommended focusing on functional processes or data flows.

Scenario 5, based on an IT specialist dialogue, highlights socio-cultural barriers in IT communication exacerbating recursions. The specialist described metaphors like “neuron” or “neocortex” as “poetry,” distinct from “fiction,” but used poetic imagery to emphasize their practical value, inadvertently mixing interpretive layers with technical reality. He asserted that thinking occurs through images, not words, citing Einstein’s visual imagery and Buddhist concepts to frame images as shaping lived reality, and insisted that metaphors embedded in functional system design (FSD) and project documentation pose no risk if developers are logical, dismissing concerns about confusion as “odd” due to overconfidence in his logic. He misinterpreted queries about joining his operational semantics (experience of a specific group) as invitations to engage with theoretical concepts, conflating the query’s intent and excluding external perspectives, reflecting cognitive biases like overconfidence and superiority bias prevalent in IT culture. These biases, including romanticizing intuitive imagery over systematic analysis and expecting ready-made tools without conceptual collaboration, perpetuate recursive ambiguities and hinder interdisciplinary integration.

The proposed system addresses these issues by separating data (query patterns, code structures) and meaning (interpretive imagery, design goals) through phenomenological reduction, resolving recursions across all scenarios. It offers a universal tool for cs.AI applications, from natural language processing (NLP) to ontology management and collaborative IT design, accessible even to non-programmers, ensuring clarity in interdisciplinary contexts like AGI research, cybernetic ontology, and IT specialist communication.

Nevertheless, the complexity of organizing such research and the need for specialized expertise limit the system’s immediate practical application. Specialists’ skepticism about formalizing subjective processes, which they view as beyond current scientific methods, exacerbated by cognitive biases like overconfidence and resistance to theoretical perspectives, poses an additional challenge. Further research, including prototype implementations, will accelerate the formalization of the language of Being, making it accessible to IT system developers. The system is particularly valuable for interdisciplinary query processing, where recursions often arise from self-reference or metaphorical ambiguity, fostering clarity and rigor in reasoning.

\section{Acknowledgments}
I express gratitude to Leonid Zhukov, Vladimir Arshinov, Yuri Garashko, Elena Churina, Andrey Kuznechenkov, Maxim Dobrokhotov, and an anonymous IT specialist for their inspiration and valuable insights during the “Philosophy + IT” seminars.

\end{document}